\documentclass[12pt]{article}
\usepackage{epsfig,amsmath,amsfonts,amssymb,amstext,afterpage,psfrag,
}
\usepackage{slashed}
\usepackage[square,comma,sort&compress,numbers]{natbib}
\allowdisplaybreaks
\newdimen\TW
\usepackage{color}
\definecolor{light-gray}{gray}{0.9}
\definecolor{dark-gray}{gray}{0.7}
\usepackage{tikz}

\long\def\symbolfootnote[#1]#2{\begingroup%
\def\thefootnote{\fnsymbol{footnote}}\footnote[#1]{#2}\endgroup}

\def\l{\langle}
\def\r{\rangle}

\def\spose#1{\hbox to 0pt{#1\hss}}
\def\lsim{\mathrel{\spose{\lower 3pt\hbox{$\mathchar"218$}}
 \raise 2.0pt\hbox{$\mathchar"13C$}}}
\def\gsim{\mathrel{\spose{\lower 3pt\hbox{$\mathchar"218$}}
 \raise 2.0pt\hbox{$\mathchar"13E$}}}

\catcode`@=11
\def\@citex[#1]#2{%
  \if@filesw\immediate\write\@auxout{\string\citation{#2}}\fi
  \def\@citea{}\@cite{\@for\@citeb:=#2\do
    {\@citea\def\@citea{,\penalty\@m}\@ifundefined
      {b@\@citeb}{{\bf ?}\@warning
{Citation `\@citeb' on page \thepage \space undefined}}%
      \hbox{\csname b@\@citeb\endcsname}}}{#1}}
\def\citer{\@ifnextchar [{\@tempswatrue\@citexr}{\@tempswafalse\@citexr[]}}
  \def\@citexr[#1]#2{%
    \if@filesw\immediate\write\@auxout{\string\citation{#2}}\fi
    \def\@citea{}\@cite{\@for\@citeb:=#2\do
      {\@citea\def\@citea{--\penalty\@m}\@ifundefined
{b@\@citeb}{{\bf ?}\@warning
{Citation `\@citeb' on page \thepage \space undefined}}%
\hbox{\csname b@\@citeb\endcsname}}}{#1}}

\begin{document}


\begin{titlepage}

\begin{flushright}
{\small
LMU-ASC~64/17\\ 
SI-HEP-2017-22\\
QFET-2017-20\\ 
IFIC/17-51\\
FTUV/17-1018\\
October 2017\\
}
\end{flushright}

\vspace{0.5cm}
\begin{center}
{\Large\bf \boldmath                                               
Complete One-Loop Renormalization of the\\ 
\vspace*{0.3cm}
Higgs-Electroweak Chiral Lagrangian 
\unboldmath}
\end{center}

\vspace{0.5cm}
\begin{center}
{\sc G. Buchalla$^a$, O. Cat\`a$^{a,b}$, A. Celis$^a$, 
M. Knecht$^c$ and C. Krause$^d$} 
\end{center}

\vspace*{0.4cm}

\begin{center}
$^a$Ludwig-Maximilians-Universit\"at M\"unchen, Fakult\"at f\"ur Physik,\\
Arnold Sommerfeld Center for Theoretical Physics, 
D-80333 M\"unchen, Germany\\
\vspace*{0.2cm}
$^b$Theoretische Physik 1, Universit\"at Siegen,\\
Walter-Flex-Stra\ss e 3, D-57068 Siegen, Germany\\
\vspace*{0.2cm}
$^c$Centre de Physique Th\'eorique (CPT),\\ 
UMR 7332 CNRS/Aix-Marseille Univ./Univ. du Sud Toulon-Var, Marseille, France\\
\vspace*{0.2cm}
$^d$IFIC, Universitat de Val\`encia -- CSIC, Apt. Correus 22085, 
E-46071 Val\`encia, Spain 
\end{center}

\vspace{1.5cm}
\begin{abstract}
\vspace{0.2cm}\noindent
Employing background-field method and super-heat-kernel expansion,
we compute the complete one-loop renormalization of the electroweak chiral
Lagrangian with a light Higgs boson. 
Earlier results from purely scalar fluctuations are confirmed 
as a special case. We also recover the one-loop renormalization of 
the conventional Standard Model in the appropriate limit. 

\end{abstract}

\vfill

\end{titlepage}

\section{Introduction}
\label{sec:intro}

Effective field theories (EFTs) for the electroweak interactions are 
nowadays part of the canonical set of techniques used at the 
LHC~\cite{deFlorian:2016spz} in the search for new physics. 
Electroweak EFTs have unique features that 
make them especially suited as discovery tools at high-energy colliders: they 
factor out in a very efficient way the known infrared physics (particle 
content and symmetries at the electroweak scale) from unknown ultraviolet 
physics. The former determine the operators of the EFT expansion, while the 
presence of the latter only affects the operator coefficients. In the context 
of the current experimental situation, where no hints of new physics have 
been detected, EFTs are extremely useful: they provide the most general, 
model-independent, unbiased parametrization of new physics compatible with 
quantum field theory requirements. 

The Higgs-electroweak chiral Lagrangian 
(HEW$\chi$L)~\cite{Feruglio:1992wf,Buchalla:2013rka,Buchalla:2013eza} 
is an effective field theory\footnote{The expression
Higgs effective field theory (HEFT) is also used by some authors.} 
of the electroweak interactions especially suited to study the 
Higgs boson nature and interactions. It is a generalization of the Standard 
Model where the Higgs boson is not required to be a weak doublet. As such, 
it is the natural upgrade of Higgs characterization schemes commonly employed 
at the LHC (most prominently the $\kappa$ formalism) into full-fledged 
quantum field theories~\cite{Buchalla:2015wfa}. 
A clear advantage of this is that radiative 
corrections can be readily implemented with known techniques, such that Higgs 
physics can be studied with increasing precision in a well-defined way.

A peculiar aspect of the HEW$\chi$L is that, as opposed to the Standard Model, 
it is nonrenormalizable: loop divergences are absorbed by counterterms, which 
introduce new operators. This is not a problem as long as those new operators 
are subleading in the EFT expansion. This can be achieved if the EFT is 
defined as a loop expansion, where the operators at a given order include the 
counterterms that absorb all the divergences up to that order. 
The paradigmatic example of such an EFT expansion is chiral perturbation theory 
(ChPT), the theory of pion dynamics~\cite{Knecht:1999ag}. 
The systematics associated with loop 
expansions has recently been revised~\cite{Buchalla:2013eza} 
and generalized power-counting 
formulas have been provided that address the specific needs of an electroweak 
EFT~\cite{Buchalla:2015wfa}. 

Power counting defines the EFT expansion and is useful to find out the 
counterterms at each loop order, but the divergence structure of the EFT can 
only be determined by the explicit renormalization of the theory at the loop 
level. While calculations based on Feynman diagrams are useful for specific 
processes~\cite{Delgado:2013hxa}, 
when it comes to the full renormalization of HEW$\chi$L they are 
rather inefficient and a path integral approach is 
preferable~\cite{Donoghue:1992dd}.
Partial results in this direction already exist in the literature, where the 
divergent structure associated with the scalar fluctuations of HEW$\chi$L has 
been worked out~\cite{Guo:2015isa,Alonso:2015fsp,Gavela:2014uta}. 

In this paper we will extend those studies and evaluate the complete one-loop 
renormalization of HEW$\chi$L. We will integrate all the one-loop fluctuations 
in the path integral using the background field method together with the 
super-heat-kernel expansion. Spe\-ci\-fi\-cally, in this paper we will 
determine the $1/\varepsilon$ poles in dimensional regularisation from 
the second 
Seeley-DeWitt coefficient. In order to do so we will re-derive a master formula 
due to 't Hooft~\cite{tHooft:1973bhk} with superspace 
methods~\cite{Neufeld:1998js,Neufeld:1998mb}, which are convenient 
when dealing with both bosonic and fermionic fluctuations. The required input 
of the master formula are the field fluctuations, which can be parametrized 
in different ways and need to be gauged-fixed. Certain choices can simplify 
the algebraic manipulations, but the final results should be independent 
of the manner the field fluctuations are parametrized. In order to cross-check 
our results, we have performed the calculation in five independent ways.

We will present our results such that the RG evolution of the coefficients of 
the NLO basis of HEW$\chi$L can be directly read off. 
We will not renormalize the 
SM parameters explicitly, albeit we provide all the ingredients to perform 
this final step. In this paper we will restrict ourselves to the formal 
aspects of the computation only. The full renormalization programme is in 
general needed when analyzing specific processes and will be carried out in 
a companion paper, with a focus on the phenomenological applications of our 
results. 

As expected, we find that fluctuations of gauge bosons and fermions define 
the renormalizable sector of the EFT and therefore do not generate new
counterterms. These stem from the nonrenormalizable sector, i.e. the pure 
scalar (Goldstone and Higgs) fluctuations,  and the mixed loops between the 
renormalizable and nonrenormalizable sectors of the theory. Our results for 
the one-loop divergences of HEW$\chi$L confirm the partial results from the 
scalar sector presented in~\cite{Guo:2015isa}. 
They also reproduce the renormalization 
of the Standard Model at one loop in the corresponding limit of parameters. 
In particular, all NLO counterterms vanish in that case. Finally, our results 
also show that chiral dimensions $d_{\chi}$, 
as defined in~\cite{Buchalla:2013eza}, are the correct expansion parameter 
for HEW$\chi$L: we consistently find that $d_{\chi}[{\cal{L}}_{\mathrm{NLO}}]=4$. 

This paper is organized as follows: in section~\ref{sec:lagrangian} 
we summarize HEW$\chi$L at leading order, mostly to set our notations. 
In section~\ref{sec:master} we discuss the generic master formula that we 
employed to calculate the one-loop divergences. Specific details for 
HEW$\chi$L are given in section~\ref{sec:details} and the final results 
for the divergences are presented in section~\ref{sec:oneloopdiv}. 
As a cross-check, in section~\ref{sec:smlimit} we show how the results 
contain the Standard Model renormalization as a particular case. 
We conclude in section~\ref{sec:concl}.

\section{Leading-order chiral Lagrangian}
\label{sec:lagrangian}

To leading order, at chiral dimension 2, the effective Lagrangian 
is given by \cite{Buchalla:2013rka,Buchalla:2013eza} 
\begin{eqnarray}\label{l2}
{\cal L} &=& -\frac{1}{2} \langle G_{\mu\nu}G^{\mu\nu}\rangle
-\frac{1}{2}\langle W_{\mu\nu}W^{\mu\nu}\rangle 
-\frac{1}{4} B_{\mu\nu}B^{\mu\nu}
+\frac{v^2}{4}\ \l L_\mu L^\mu \r\, F(h) +
\frac{1}{2} \partial_\mu h \partial^\mu h - V(h)
\nonumber\\
&& +\bar\psi i\!\not\!\! D\psi - \bar\psi m(h,U)\psi
\end{eqnarray}
$G$, $W$ and $B$ are the gauge fields of $SU(3)_C$, $SU(2)_L$ and
$U(1)_Y$, respectively. The trace is denoted by $\langle\ldots\rangle$.
$h$ is the Higgs field. The electroweak Goldstone bosons are parametrized as
$U=\exp(2i\varphi/v)$, where $\varphi=\varphi^a T^a$. $T^a$ are the
generators of $SU(2)$, normalized as $\langle T^a T^b\rangle=\delta^{ab}/2$,
and $v=246\,{\rm GeV}$ is the electroweak scale.  
We define  
\begin{equation}\label{dcovu}
L_\mu=i UD_\mu U^\dagger\, , \quad {\rm where}
\quad D_\mu U=\partial_\mu U + i g W_\mu U -i g' B_\mu U T_3\, ,
\qquad \tau_L=U T_3 U^\dagger
\end{equation}

The Standard-Model (SM) fermions are collected in the field 
$\psi=(u_i,d_i,\nu_i,e_i)^T$. Here $i$ is the generation index, $u_i$ and $d_i$ 
are color triplets, and the $u_i,d_i,\nu_i$ and $e_i$ are Dirac spinors.
The covariant derivative is
\begin{equation}\label{dcovpsi}
D_\mu \psi=\left(\partial_\mu +i g_s G_\mu + i g W_\mu P_L +
i g' B_\mu (Y_L P_L + Y_R P_R)\right)\psi 
\end{equation}
$P_L$, $P_R$ are the left and right chiral projectors.
Weak hypercharge is described by the diagonal matrices
\begin{equation}
Y_L={\rm diag}(1/6,1/6,-1/2,-1/2)\, ,\qquad
Y_R={\rm diag}(2/3,-1/3,0,-1)
\end{equation}
The Yukawa term can be compactly expressed as
\begin{equation}\label{myukawa}
m(h,U)\equiv U {\cal M}(h) P_R +{\cal M}^\dagger(h) U^\dagger P_L
\end{equation}
with ${\cal M}$ the block-diagonal mass matrix, acting on $\psi$,
\begin{equation}\label{mmdef}
{\cal M}={\rm diag}({\cal M}_u,{\cal M}_d,{\cal M}_\nu,{\cal M}_e) 
\end{equation}
The entries ${\cal M}_f\equiv{\cal M}_f(h)$ 
are matrices in generation space and functions of $h$.
It is understood that the right-handed neutrinos are absent when we
assume SM particle content. Accordingly, we will take
${\cal M}_\nu=0$ in our calculation.

The Higgs-dependent functions can be expanded as
\begin{equation}\label{fvdef}
F(h)=1+\sum^\infty_{n=1} F_{n} \left(\frac{h}{v}\right)^n\, ,\qquad 
V(h)=v^4\sum^\infty_{n=2} V_{n} \left(\frac{h}{v}\right)^n\, ,\qquad
{\cal M}(h)=\sum^\infty_{n=0} {\cal M}_{n} \left(\frac{h}{v}\right)^n
\end{equation}

\section{Master formula for one-loop divergences}
\label{sec:master}

In this section, we review the master formula giving the
one-loop divergences of a general Lagrangian including
both spin 0 and spin 1 bosons, as well as fermions.
An equivalent result has been obtained a long time ago in
\cite{tHooft:1973bhk} and in \cite{Jack:1984vj,Lee:1984ud}.
We re-derive it here using the super-heat-kernel framework
of \cite{Neufeld:1998js}. The formula will be the basis
for the calculation of the one-loop renormalization of the
electroweak chiral Lagrangian in (\ref{l2}).
The discussion will also serve to fix our notation.

Starting from our general Lagrangian, we expand each
field around a classical background configuration.
The fluctuating parts of the various fields are
denoted generically as $\xi$, $\omega_\mu$, and
$\chi$ for the spin 0, spin 1 and spin 1/2 Dirac fields,
respectively. Notice that all internal indices have
been omitted, so that these fields denote in general
multi-component objects. The bosonic fields are furthermore
conveniently collected in a single multi-component object
\begin{equation}
\phi_i = (\xi, \omega_\mu),\quad \phi^i = (\xi , -\omega^\mu )
.
\end{equation}
Assuming that the Lagrangian we started with is at most bilinear in the
fermion fields, the part that is quadratic in the fluctuations takes,
up to an irrelevant total derivative, the general form
\begin{equation}\label{lfluct}
{\cal L}_2= -\frac{1}{2}\phi^i A_i^{\,j} \phi_j
+ \bar\chi\left( i\!\not\! \partial - G\right) \chi
+\bar\chi\Gamma^i\phi_i +\phi^i\bar\Gamma_i\chi
\end{equation}
with
\begin{equation}
A = (\partial^\mu + N^\mu) (\partial_\mu + N_\mu) + Y
\end{equation}
For the fluctuating gauge fields, the Feynman gauge has to be used to ensure
the canonical form of the kinetic term for the bosons in (\ref{lfluct}).
The Dirac matrix $G$ can be written as
\begin{equation}\label{gdef}
G\equiv (r+ \rho_\mu \gamma^\mu)P_R + (l+ \lambda_\mu \gamma^\mu)P_L
\end{equation}
The quantities $Y$, $N$, $r$, $l$, $\rho$, $\lambda$ are bosonic,
while $\Gamma$ and $\bar\Gamma$ are Dirac spinors.
They all depend on the background fields. Notice that in
all generality one could also add a tensor contribution
$\sigma_{\mu\nu} t^{\mu\nu}$ to $G$. Since such a term does
not arise in the case we will study, we do not consider it.

The Gaussian integral over the bosonic and fermionic variables in
\begin{equation}\label{seffdef}
e^{iS_{\rm eff}} \sim \int [d\phi_i\, d\chi\, d\bar\chi]
e^{i\int d^Dx\, {\cal L}_2(\phi_i,\chi,\bar\chi)}
\end{equation}
(gauge -fixing in the way described below is understood)
leads to an expression for $S_{\rm eff}$ in terms of the fluctuation
operator
in ${\cal L}_2$. Keeping only the terms needed for the divergent part
of $S_{\rm eff}$, this expression can be written as \cite{Neufeld:1998js}
\begin{equation}\label{seffstr}
S_{\rm eff}=\frac{i}{2} {\rm Str}\, \ln\Delta
\end{equation}
where
\begin{equation}\label{deltadef}
\Delta\equiv\left(
\begin{array}{cc}
A & \sqrt{2}\bar\Gamma\gamma_5 B\gamma_5 \\
-\sqrt{2}\Gamma & B\gamma_5 B\gamma_5
\end{array}
\right) ,
\qquad B\equiv i\!\not\! \partial - G
\end{equation}
Here the supertrace ${\rm str}$ of a general supermatrix
\begin{equation}\label{mabdef}
M =\left(
\begin{array}{cc}
a & \alpha \\
\beta & b
\end{array}
\right)
\end{equation}
with $a$, $b$ bosonic and $\alpha$, $\beta$ fermionic sub-matrices,
is defined by
\begin{equation}\label{strdef}
{\rm str}\, M = {\rm tr}\, a - {\rm tr}\, b
\end{equation}
The analogous trace operations ${\rm Str}$ and ${\rm Tr}$ include
an integration over space-time.

The operator $\Delta$ in (\ref{deltadef}) has the canonical form
\begin{equation}\label{deltacan}
\Delta \equiv (\partial^\mu +\Lambda^\mu)(\partial_\mu +\Lambda_\mu)
+\Sigma
\end{equation}
which defines the supermatrices $\Lambda_\mu$ and $\Sigma$.
In Euclidian space, the differential operator $\Delta$ is elliptic,
and the divergent part, in four dimensions, of the effective action
is given by the second Seeley-DeWitt coefficient of the
corresponding heat-kernel expansion.
The computation of the second Seeley-DeWitt coefficient
for an operator like $\Delta$ is described in \cite{Neufeld:1998js}.
The divergent part of the dimensionally regularized one-loop 
effective Lagrangian is then given, in Minkowski space, by
\begin{equation}\label{ldiv}
{\cal L}_{\rm div} =\frac{1}{32\pi^2\varepsilon}\,
{\rm str}\, \left[\frac{1}{12}\Lambda^{\mu\nu}\Lambda_{\mu\nu}
+\frac{1}{2}\Sigma^2\right]
\end{equation}
where $\varepsilon = 2 - d/2$ and
\begin{equation}\label{lmunudef}
\Lambda_{\mu\nu}\equiv \partial_\mu \Lambda_\nu -\partial_\nu \Lambda_\mu
+[\Lambda_\mu,\Lambda_\nu]
\end{equation}

Extracting $\Lambda_\mu$ and $\Sigma$ from (\ref{deltadef}) and
(\ref{deltacan}), and performing the traces over Dirac matrices
explicitly \cite{Borodulin:2017pwh,Mertig:1990an,Jamin:1991dp,Shtabovenko:2016sxi,mathematica},
one finally arrives at the master formula \cite{tHooft:1973bhk}
\begin{eqnarray}\label{masterform}
{\cal L}_{\rm div} = \frac{1}{32\pi^2\varepsilon}\Bigg( &&{\rm tr}  \Bigg[
\frac{1}{12} N^{\mu\nu}N_{\mu\nu}+\frac{1}{2} Y^2
-\frac{1}{3}\left(\lambda^{\mu\nu}\lambda_{\mu\nu}+\rho^{\mu\nu}\rho_{\mu\nu}\right)
+ 2 D^\mu l D_\mu r - 2 lrlr \Bigg]\nonumber \\
&& + \bar\Gamma\left( i\!\not\! \partial +i\!\not\!\! N +
\frac{1}{2}\gamma^\mu G\gamma_\mu\right)\Gamma\Bigg)
\end{eqnarray}
with
\begin{equation}\label{nydef}
N_{\mu\nu} \equiv  \partial_\mu N_\nu -\partial_\nu N_\mu +[N_\mu, N_\nu]
\end{equation}
\begin{equation}\label{lrmunudef}
\lambda_{\mu\nu} \equiv  \partial_\mu \lambda_\nu -\partial_\nu
\lambda_\mu  +
i [\lambda_\mu, \lambda_\nu]\, ,
\qquad
\rho_{\mu\nu} \equiv  \partial_\mu \rho_\nu -\partial_\nu \rho_\mu  +
i [\rho_\mu, \rho_\nu]
\end{equation}
\begin{equation}\label{dldrdef}
D_\mu l \equiv  \partial_\mu l + i \rho_\mu l - i l \lambda_\mu\, , \qquad
D_\mu r \equiv  \partial_\mu r + i \lambda_\mu r - i r \rho_\mu
\end{equation}
In (\ref{masterform}) the terms bilinear in $N_{\mu\nu}$, $Y$, in
$\lambda_{\mu\nu}$, $\rho_{\mu\nu}$, $l$, $r$, and in $\Gamma$,
$\bar\Gamma$,
originate, respectively, from pure bosonic loops, pure fermionic loops,
and mixed contributions with both bosons and fermions in the loop.
The expression in (\ref{masterform}) holds up to surface terms that
we have dropped. The ghost contribution for non-abelian gauge fields
has to be added and will be discussed in section \ref{sec:details}. 

\section{Technical aspects of the calculation}
\label{sec:details}

In order to implement the background field method \cite{Abbott:1981ke}, 
all fields are split additively into background and quantum components 
except for the Goldstone boson matrix $U$, which is expanded in 
multiplicative form following~\cite{Dittmaier:1995cr,Dittmaier:1995ee}.    
This allows us to remove the background Goldstone fields from the Lagrangian 
using a generalization of the St\"uckelberg formalism for the background 
field method~\cite{Cheyette:1987qz}.   The fact that no background Goldstone 
fields are present simplifies intermediate steps of the calculation.     
The background Goldstone fields are then recovered at the end of the 
calculation by inverting the St\"uckelberg 
transformation~\cite{Dittmaier:1995cr,Dittmaier:1995ee}.

In the presence of non-abelian gauge fields, one needs to add the 
contribution arising from ghost loops. 
Let us denote the fluctuating components of the $B_\mu$, $W_\mu$ and
$G_\mu$ fields by $b_\mu$, $\omega_\mu$ and $\varepsilon_\mu$,
respectively. Choosing background-covariant gauge conditions
\begin{equation}
{\cal L}_{\rm g.f.} = 
-\frac{1}{2} (\partial^\mu b_\mu)^2 - \langle (D^\mu \omega_\mu)^2 \rangle
- \langle (D^\mu \varepsilon_\mu)^2 \rangle
\end{equation}
the additional contribution to the divergent part of the one-loop effective
Lagrangian reads
\begin{equation}
{\cal L}_{\rm div ; ghost}=\frac{1}{32\pi^2\varepsilon}\Bigg(
\frac{1}{3} g^2 C_2^W \langle W_{\mu\nu}W^{\mu\nu} \rangle +
\frac{1}{3} g_s^2 C_2^G \langle G_{\mu\nu}G^{\mu\nu} \rangle
\Bigg)
\end{equation}
where $C_2^W = 2$ and $C_2^G= 3$ are the quadratic Casimirs for the $SU(2)$ 
and $SU(3)$ gauge groups, respectively. 

Though, within Feynman gauge, the total result for the one-loop divergences 
will be independent of the gauge fixing choice (up to field redefinitions), 
the individual contributions to the master formula 
will depend in general on the gauge fixing term. We will report our 
results using the (electroweak) gauge fixing term
\begin{align}  \label{gfix2}
{\cal L}_{\rm g.f.} = 
-\frac{1}{2} \left(\partial^\mu b_\mu +\frac{g' v}{2} F\varphi_3\right)^2 
-\frac{1}{2} \left(D^\mu \omega^a_\mu-\frac{gv}{2} F \varphi^a\right)^2 
\end{align}
with $\varphi^a$ the fluctuating Goldstone fields.
This gives the following divergent contribution to the one-loop effective 
Lagrangian from the ghost sector
\begin{align}
{\cal L}_{\rm div ; ghost}=\frac{1}{32\pi^2\varepsilon}\Bigg(
\frac{2}{3} g^2 \langle W_{\mu\nu}W^{\mu\nu} \rangle -
(3g^4 + 2 g^2 g'^2 + g'^4)\frac{v^4}{16} F^2 \Bigg)
\end{align}
The gauge fixing term \eqref{gfix2} is invariant under background-gauge 
transformations and cancels the mixing between the Goldstone fields and the 
gauge fields arising from the Goldstone-boson kinetic term.  A similar 
choice of gauge fixing was used in~\cite{Dittmaier:1995ee}.  In our 
calculation we checked explicitly that, as expected, the total result for 
the divergent contributions to the one-loop effective Lagrangian is the same 
with the two gauge fixing terms specified above.

\section{One-loop divergences}
\label{sec:oneloopdiv}

In this section we give the explicit expressions for the
complete one-loop divergences of the Higgs-electroweak chiral Lagrangian.
They provide the counterterms that renormalize the theory at this order.
These formulas are the main result of our paper.
The divergences can be separated into the contributions
from the electroweak sector and those from QCD, which we present in turn.

\subsection{Electroweak sector}
\label{sec:ewdiv}

We define
\begin{equation}\label{kapdef}
\eta\equiv\frac{h}{v}\, ,\qquad
\kappa\equiv \frac{1}{2}F' F^{-1/2}\, ,\qquad 
{\cal B}\equiv -F^{-1/2}\kappa'=\frac{F'^2}{4 F^2}-\frac{F''}{2F}
\end{equation}
Here and in the following, a prime on $\eta$-dependent functions 
denotes differentiation with respect to this variable.

For the contributions to the master formula (\ref{masterform}) 
we finally obtain

\begin{eqnarray}
&&{\rm tr}\left(\frac{1}{12}N^{\mu\nu} N_{\mu\nu} +\frac{1}{2} Y^2\right) =
\frac{22}{3} g^2 \langle W^{\mu\nu}W_{\mu\nu}\rangle\nonumber\\
&&-(g'^2 + g^2(\kappa^2+2))\frac{v^2}{2}F\,\langle L^\mu L_\mu\rangle
+g'^2 v^2 (1-\kappa^2)F\,\langle\tau_LL^\mu\rangle
\langle\tau_LL_\mu\rangle\nonumber\\
&&-(3 g^2 + g'^2)v^2\kappa^2\, \partial^\mu\eta \partial_\mu\eta
+(3 g^4 + 2 g^2 g'^2 + g'^4)\frac{v^4}{16} F^2   \label{n2y2a}\\   
&&+(3 g^4 + 2 g^2 g'^2 + g'^4)\frac{v^4}{32}F^2+\frac{3 g^2 + g'^2}{8}F'V'+
\frac{3}{8}\left(\frac{F'V'}{F v^2}\right)^2+
\frac{1}{2}\left(\frac{V''}{v^2}\right)^2\nonumber\\
&&+\left((3g^2+g'^2)\frac{v^2}{4}F+\frac{3}{2}\frac{F'V'}{F v^2}\right){\cal B}
\, \partial^\mu\eta \partial_\mu\eta\nonumber\\
&&-\left[(\kappa^2-1)\left((2g^2+g'^2)\frac{v^2}{8}F+\frac{F'V'}{2Fv^2}\right)
-\frac{V'' F}{2v^2}{\cal B}\right]\, \langle L^\mu L_\mu\rangle\nonumber\\
&&+\left((3g^2+g'^2)\frac{v^2}{4} F+\frac{3}{2}\frac{F'V'}{F v^2}\right)
\frac{F^{-1}}{v^2}
\left(
\bar\psi_LU\left(\frac{F'}{2}{\cal M}'-{\cal M}\right)\psi_R +{\rm h.c.}
\right)\nonumber\\
&&+\frac{V''}{v^4}\left(\bar\psi_LU {\cal M}'' \psi_R +{\rm h.c.}\right)
-\frac{\kappa^2+1}{24}\left(2g^2\langle W^{\mu\nu}W_{\mu\nu}\rangle
+g'^2 B^{\mu\nu} B_{\mu\nu}\right) \label{n2y2b}\\ 
&&+\frac{\kappa^2-1}{6} gg' \langle\tau_L W^{\mu\nu}\rangle B_{\mu\nu}
-\frac{\kappa^2-1}{12}\left(ig\langle W^{\mu\nu}[L_\mu,L_\nu]\rangle
+ig' B^{\mu\nu}\langle\tau_L [L_\mu,L_\nu]\rangle\right)\nonumber\\
&&-\frac{\kappa\kappa'}{3}\partial^\mu\eta\left(g\langle W_{\mu\nu}L^\nu\rangle
-g' B_{\mu\nu}\langle\tau_L L^\nu\rangle\right)
+\frac{1}{4}g'^2 v^2(\kappa^2-1)F
\langle\tau_L L^\mu\rangle\langle\tau_L L_\mu\rangle\nonumber\\
&&+\frac{(\kappa^2-1)^2}{6} \langle L_\mu L_\nu\rangle^2
+\left(\frac{(\kappa^2-1)^2}{12}+\frac{F^2{\cal B}^2}{8}\right)
\langle L^\mu L_\mu\rangle^2 
+\frac{2}{3}\kappa'^2\langle L_\mu L_\nu\rangle \partial^\mu\eta\partial^\nu\eta
\nonumber\\
&&-\left((\kappa^2-1){\cal B}+\frac{\kappa'^2}{6}\right)
\langle L^\mu L_\mu\rangle \partial^\nu\eta\partial_\nu\eta
+\frac{3}{2}{\cal B}^2(\partial^\mu\eta\partial_\mu\eta)^2\nonumber\\
&&+\langle L^\mu L_\mu\rangle
\left[\frac{F{\cal B}}{2v^2}\bar\psi_L U{\cal M}''\psi_R
-\frac{\kappa^2-1}{Fv^2}
\bar\psi_L U\left(\frac{F'}{2}{\cal M}'-{\cal M}\right)\psi_R +{\rm h.c.}
\right]\nonumber\\
&&+\frac{2\kappa'}{v^2}\partial^\mu\eta\left(
i\bar\psi_L L_\mu U(F^{-1/2}{\cal M})'\psi_R +{\rm h.c.}\right)\nonumber\\
&&+\frac{3{\cal B}}{Fv^2} \partial^\mu\eta\partial_\mu\eta\left(\bar\psi_L U
\left(\frac{F'}{2}{\cal M}'-{\cal M}\right)\psi_R +{\rm h.c.}\right)
\nonumber\\
&&+\frac{3F^{-2}}{2v^4}\left(\bar\psi_L U
\left(\frac{F'}{2}{\cal M}'-{\cal M}\right)\psi_R +{\rm h.c.}\right)^2
+\frac{1}{2v^4}\left(\bar\psi_L U{\cal M}''\psi_R +{\rm h.c.}\right)^2
\nonumber\\
&&+\frac{4}{v^4}\left(i\bar\psi_L U T^a
\left(F^{-1/2} {\cal M}\right)'\psi_R +{\rm h.c.}\right)^2  \label{n2y2c}
\end{eqnarray}

The terms in (\ref{n2y2a}) arise from loops with gauge fields
and include the ghost contribution.
The remaining ones come from loops with scalars.
Operators in (\ref{n2y2b}) have the form of terms in the leading-order
Lagrangian, with the exception of 
$\langle\tau_L L^\mu\rangle\langle\tau_L L_\mu\rangle$ and the gauge kinetic 
terms $\langle W^{\mu\nu}W_{\mu\nu}\rangle$, $B^{\mu\nu}B_{\mu\nu}$
multiplied by powers of the Higgs field $h^n$, $n\geq 1$.
All the operators in (\ref{n2y2c}) arise only at next-to-leading order.

\begin{eqnarray}\label{lrmunu}
&& -\frac{1}{3}\, {\rm tr}(\lambda^{\mu\nu}\lambda_{\mu\nu} +
  \rho^{\mu\nu}\rho_{\mu\nu}) = \nonumber\\
&& -\frac{1}{2}\langle W^{\mu\nu}W_{\mu\nu}\rangle\, \frac{2}{3}(N_c+1)f g^2
 -\frac{1}{4} B^{\mu\nu} B_{\mu\nu}\left(\frac{22 N_c}{27} + 2\right) f g'^2
\end{eqnarray}
where $N_c$ is the number of colors and $f$ the number of 
fermion generations.

\begin{eqnarray}\label{dldr}
&& 2\,  {\rm tr}\left(D^\mu l D_\mu r -  lrlr\right) =\\
&& 2\langle\partial^\mu{\cal M}^\dagger \partial_\mu{\cal M}\rangle +
4i\langle(\partial^\mu{\cal M}^\dagger{\cal M}-
  {\cal M}^\dagger \partial^\mu{\cal M})T_3\rangle \langle\tau_L L_\mu\rangle
 +\langle{\cal M}^\dagger {\cal M}\rangle \langle L^\mu L_\mu\rangle
 - 2 \langle ({\cal M}^\dagger {\cal M})^2\rangle \nonumber
\end{eqnarray}

\begin{eqnarray}\label{gammabargamma}
&&\bar\Gamma\left( i\!\not\! \partial +i\!\not\!\! N +
\frac{1}{2}\gamma^\mu G\gamma_\mu\right)\Gamma =
\frac{4}{v^2}\bar\psi_LUT^a{\cal M}F^{-1/2} i\!\not\! \partial
\left({\cal M}^\dagger F^{-1/2}\right)T^a U^\dagger\psi_L\nonumber\\
&&+\frac{4}{v^2}\bar\psi_L UT^a{\cal M}{\cal M}^\dagger F^{-1} 
   T^a U^\dagger i\!\not\!\! D\psi_L 
+\frac{1}{v^2}\bar\psi_L \not\!\! L 
  U{\cal M}{\cal M}^\dagger U^\dagger F^{-1}\psi_L \nonumber\\
&&+\frac{1}{v^2}\bar\psi_L U{\cal M}' i\!\not\! \partial{\cal M}'^\dagger 
    U^\dagger \psi_L 
+\frac{1}{v^2}\bar\psi_L U{\cal M}'{\cal M}'^\dagger U^\dagger 
  (i\!\not\!\! D +\not\!\! L) \psi_L
\nonumber\\
&&-\frac{\kappa}{v^2}F^{-1/2}\left(\bar\psi_L U {\cal M}'{\cal M}^\dagger 
   U^\dagger\not\!\! L\psi_L +{\rm h.c.}\right) \nonumber\\
&&+\frac{3}{v^2}\bar\psi_R{\cal M}^\dagger F^{-1/2} 
  i\!\not\!\! D\left({\cal M}F^{-1/2}\psi_R\right)
+\frac{1}{v^2}\bar\psi_R{\cal M}'^\dagger 
  i\!\not\!\! D\left({\cal M}'\psi_R\right)\nonumber\\
&&-\frac{F^{-1}}{v^2}\bar\psi_R{\cal M}^\dagger 
 U^\dagger \not\!\! L U{\cal M}\psi_R 
-\frac{1}{v^2}\bar\psi_R{\cal M}'^\dagger 
 U^\dagger \not\!\! L U{\cal M}'\psi_R\nonumber\\ 
&&+\frac{\kappa}{v^2}F^{-1/2}\left(\bar\psi_R{\cal M}^\dagger U^\dagger
   \not\!\! L U{\cal M}'\psi_R +{\rm h.c.}\right)\nonumber\\
&&-\frac{8}{v^2}F^{-1}\left(\bar\psi_L U T^a{\cal M}{\cal M}^\dagger T^a
   {\cal M}\psi_R +{\rm h.c.}\right)
+\frac{2}{v^2}\left(\bar\psi_L U{\cal M}'{\cal M}^\dagger{\cal M}'\psi_R
   +{\rm h.c.}\right)\nonumber\\
&&+\bar\psi_L\left(\frac{3}{2}g^2+2 g'^2 Y^2_L\right)i\!\not\!\! D\psi_L
+\bar\psi_R\, 2g'^2 Y^2_R i\!\not\!\! D\psi_R
-8 g'^2\left(\bar\psi_L Y_L U{\cal M}Y_R\psi_R +{\rm h.c.}\right)
\end{eqnarray}

\subsection{QCD sector}
\label{sec:qcddiv}

At one-loop order, QCD and electroweak renormalization 
can be treated separately. 
To obtain the one-loop divergences from QCD, we expand the
Lagrangian in (\ref{l2}) to second order in fluctuations of
the quark and gluon fields, treating gauge fixing and ghosts
in the usual way. We follow again the procedure outlined in 
section \ref{sec:master}. For the divergent part of the
one-loop effective Lagrangian we find
\begin{equation}\label{ldivqcd1}
{\cal L}_{\rm div,QCD}\equiv \frac{1}{32\pi^2\varepsilon}  L_{\rm div,QCD}
\end{equation}
\begin{equation}\label{ldivqcd2}
L_{\rm div,QCD}=\frac{22 N_c - 4 N_f}{6}g^2_s \langle G^{\mu\nu}G_{\mu\nu}\rangle 
+ 2 g^2_s C_F \, \bar q\left(i\!\not\!\! D 
  -4(U{\cal M}_q P_R+{\cal M}^\dagger_q U^\dagger P_L)\right)q
\end{equation}
Here $C_F=(N^2_c-1)/(2N_c)$ and $N_f$ is the number of quark flavors.
In analogy to section \ref{sec:lagrangian} we take
$q=(u,d)^T$ and ${\cal M}_q={\rm diag}({\cal M}_u,{\cal M}_d)$.

\subsection{Renormalization}
\label{sec:renorm}

The divergences displayed in eq. (\ref{n2y2c})
are absorbed by the effective Lagrangian at NLO, whose structure 
has been systematically analysed in \cite{Buchalla:2013rka},
\begin{equation}
{\cal L}_{\rm NLO} = \sum_i \frac{v^{6-d_i}}{\Lambda^2} F_i (h) {\cal O}_i ,
\end{equation}
with $\Lambda = 4 \pi v$. A complete basis of 
operators ${\cal O}_i$ is also provided in \cite{Buchalla:2013rka}. 
Upon minimal subtraction of the one-loop divergences displayed in 
(\ref{n2y2c}), 
the functions $F_i(h)$ will depend on the renormalization scale $\mu$, with
\begin{equation}
F_i (h ; \mu)  = F_i (h ; \mu_0) + \beta_i(h) \ln(\mu/\mu_0) .
\end{equation}
As announced in the Introduction, we will not give the explicit
expressions of the beta functions $\beta_i(h)$ corresponding to
the complete basis of operators ${\cal O}_i$ here, leaving this
last step for future work. At this stage, let us just make a remark
concerning the divergences given in eqs. (\ref{n2y2a}), (\ref{n2y2b}),
which correspond
to terms already present in the lowest-order effective Lagrangian
${\cal L}$ in (\ref{l2}). The form of the latter is the most general up
to field redefinitions of $h(x)$. The latter have been used in order to 
(see, for instance, the discussion in appendix A of \cite{Buchalla:2013rka}):

\begin{description}
\item 
i) remove any arbitrary functions of $h$ in front of
the kinetic terms of the Higgs field and of the fermion fields;
\item 
ii) remove a linear term in the Higgs potential, i.e. imposing
$V^\prime (0)=0$.
\end{description}

These features are modified by the structure of the one-loop divergences
given in (\ref{n2y2a}), (\ref{n2y2b}). 
One thus needs to perform the appropriate field redefinition
of $h(x)$ in order to restore them.

\section{Standard-Model limit}
\label{sec:smlimit}

The Higgs dynamics in the chiral Lagrangian (\ref{l2}) is encoded
in the functions $F(h)$, $V(h)$ and ${\cal M}(h)$.
The renormalizable Standard Model (SM) is recovered in the limit 
($\eta\equiv h/v$)
\begin{equation}\label{smlimit}
F = (1+\eta)^2\, ,\qquad
V = \frac{m^2_h v^2}{8}\left( -2 (1+\eta)^2 + (1+\eta)^4 \right)\, ,\qquad
{\cal M} = {\cal M}_0\, (1+\eta)
\end{equation}
In this limit, all nonrenormalizable operators disappear from
the divergent part of the effective Lagrangian given in
section \ref{sec:oneloopdiv}. The remaining expressions give
the one-loop divergences of the Standard Model, from which
the well-known one-loop beta functions of the SM couplings can be obtained. 
We find agreement with the beta functions compiled in \cite{Celis:2017hod}.  
This serves as an important check of our results. 

It will be useful to write the scalar fields in terms of the usual
Higgs doublet $\Phi$ and $\tilde\Phi$, where 
$\tilde\Phi_i=\epsilon_{ij}\Phi^*_j$.
The relation to the chiral coordinates is given by
\begin{equation}\label{phihu}
(\tilde\Phi,\Phi)=\frac{v}{\sqrt{2}}(1+\eta) U
\end{equation}
and we have
\begin{equation}\label{phidagphi}
D^\mu\Phi^\dagger D_\mu\Phi =\frac{1}{2}\partial^\mu h\partial_\mu h
+\frac{v^2}{4}\langle L^\mu L_\mu\rangle (1+\eta)^2\, ,\qquad
\Phi^\dagger\Phi=\frac{v^2}{2}(1+\eta)^2
\end{equation}
We will also use the SM relations
\begin{equation}\label{smrel}
M^2_W=\frac{1}{4}g^2 v^2\, ,\quad M^2_Z=\frac{1}{4}(g^2 + g'^2)v^2\, ,\quad
m^2_h=2\mu^2=\lambda v^2\, ,\quad m^2_t=\frac{1}{2}y^2_t v^2
\end{equation}
This defines the parameters $\mu^2$ and $\lambda$ of the Higgs potential 
and the top-quark Yukawa coupling $y_t$.
In general, the Yukawa matrix ${\cal Y}$ and the mass matrix ${\cal M}_0$
are related through
\begin{equation}\label{ymrel}
 {\cal M}_0=\frac{v}{\sqrt{2}}{\cal Y}\, ,\qquad {\rm where}\quad 
{\cal Y}={\rm diag}({\cal Y}_u,{\cal Y}_d,{\cal Y}_\nu,{\cal Y}_e)
\end{equation}
such that
\begin{equation}\label{umphiy}
U{\cal M}_0 (1+\eta) = (\tilde\Phi,\Phi)\, {\cal Y}
\end{equation}
and we have
\begin{equation}\label{tracemm}
\langle{\cal M}_0^\dagger {\cal M}_0\rangle =
N_c(m^2_t + m^2_c + m^2_u + m^2_b + m^2_s + m^2_d)+ m^2_\tau + m^2_\mu + m^2_e
\approx N_c m^2_t
\end{equation}
Here the trace is over color, family and isospin indices, 
and includes quarks and leptons. Below we will sometimes retain only
the top-quark part to simplify expressions.
Similarly, 
$\langle ({\cal M}_0^\dagger {\cal M}_0)^2\rangle \approx N_c m^4_t$.

\subsection{Electroweak sector -- bosonic part}

We start with the bosonic part of the electroweak sector
collected in (\ref{n2y2a}) -- (\ref{dldr}).
Using the relations above we find in the SM limit
\begin{eqnarray}
&& 32\pi^2\varepsilon\, {\cal L}^{\rm SM}_{\rm div,EWb} =\nonumber\\
&& -\frac{1}{2}\langle W^{\mu\nu}W_{\mu\nu}\rangle
\left(-\frac{44}{3}+\frac{2}{3}(N_c+1)f+\frac{1}{3}\right)g^2
-\frac{1}{4}B^{\mu\nu}B_{\mu\nu}
\left(\left(\frac{22 N_c}{27}+2\right)f+\frac{1}{3}\right)g'^2
\nonumber\\
&& +D^\mu\Phi^\dagger D_\mu\Phi\left(-6g^2-2 g'^2+2N_c y^2_t\right)
+\mu^2\Phi^\dagger\Phi\left(-\frac{3}{2}g^2-\frac{1}{2}g'^2-6\lambda\right)
\nonumber\\
&&-\frac{\lambda}{2}(\Phi^\dagger\Phi)^2
\left(-3g^2-g'^2-12\lambda-\frac{3}{4\lambda}(3 g^4+2 g^2 g'^2 + g'^4)
+\frac{4 N_c}{\lambda} y^4_t\right)\label{lsmdivewb}
\end{eqnarray}
For the gauge-kinetic terms the contributions from gauge fields,
fermions ($\sim f$) and scalars ($+1/3$) have been written separately.

Renormalizing fields 
\begin{equation}\label{zfields}
W^\mu_{(0)}=W^\mu Z^{1/2}_W\, ,\quad B^\mu_{(0)}=B^\mu Z^{1/2}_B\, ,\quad
\Phi_{(0)}=\Phi Z^{1/2}_\Phi
\end{equation}
and couplings 
\begin{equation}\label{zcoupl}
g_{(0)}=g\, \nu^\varepsilon Z_g\, ,\quad 
g'_{(0)}=g'\, \nu^\varepsilon Z_{g'}\, ,\quad
\lambda_{(0)}=\lambda\, \nu^{2\varepsilon} Z_\lambda\, ,\quad 
\mu^2_{(0)}=\mu^2 Z_m
\end{equation}
where $\nu$ is a renormalization scale, we obtain from the leading-order
SM Lagrangian the counter-terms  
\begin{eqnarray}\label{lsmctewb}
{\cal L}^{\rm SM}_{\rm CT,EWb} &=& 
-\frac{1}{2}\langle W^{\mu\nu} W_{\mu\nu}\rangle (Z_W-1)
-\frac{1}{4} B^{\mu\nu} B_{\mu\nu} (Z_B-1)
+D^\mu\Phi^\dagger D_\mu\Phi (Z_\phi-1)\nonumber\\
&&+\mu^2 \Phi^\dagger\Phi (Z_\Phi Z_m -1)
-\frac{\lambda}{2} (\Phi^\dagger\Phi)^2 (Z^2_\Phi Z_\lambda -1)
\end{eqnarray}
In the background field method we are using, the renormalization factors 
of gauge couplings and gauge fields are related through
\begin{equation}\label{zgzw}
Z_g=Z^{-1/2}_W\, ,\qquad Z_{g'}=Z^{-1/2}_B
\end{equation}

To one-loop order the renormalization factor $Z_X$ for quantity $X$
can be written as
\begin{equation}\label{zxax}
Z_X=1+\frac{A_X}{16\pi^2\varepsilon}
\end{equation} 
where $A_X$ only depends on couplings. 
The beta function for parameter $X$, defined by \cite{Celis:2017hod}
\begin{equation}\label{betadef}
\frac{dX}{dt}\equiv \frac{1}{16\pi^2}\beta_X\, , \qquad t=\ln\nu
\end{equation}
is then given by
\begin{equation}\label{betaxax}
\beta_X=-X\, \frac{1}{\varepsilon}\frac{dA_X}{dt}
\end{equation}
and can be evaluated using
\begin{equation}\label{dgndt}
\frac{d}{dt}\left( g^{n_g} g'^{n_{g'}} \lambda^{n_\lambda} y^{n_y}_t\right)
=-(n_g + n_{g'} + 2 n_\lambda + n_y)\varepsilon\, \,
g^{n_g} g'^{n_{g'}} \lambda^{n_\lambda} y^{n_y}_t + 
{\cal O}\left(\frac{1}{16\pi^2}\right)
\end{equation}
for the scaling of products of couplings in dimensional regularization.

Requiring that the $1/\varepsilon$ poles of ${\cal L}^{\rm SM}_{\rm div,EWb}$
in (\ref{lsmdivewb})  are cancelled
by adding the coun\-ter\-terms in (\ref{lsmctewb}) fixes the
renormalization factors $Z_\Phi$, $Z_g$, $Z_{g'}$, $Z_\lambda$ and $Z_m$.
Using (\ref{zxax}), (\ref{betaxax}) and (\ref{dgndt}), we recover the one-loop 
beta functions of the SM couplings $g$, $g'$, $\lambda$ and $\mu^2$:
\begin{eqnarray}
\beta_g &=& -\left(\frac{22}{3}-\frac{N_c+1}{3}f-\frac{1}{6}\right)g^3
=-\frac{19}{6}g^3\\
\beta_{g'} &=& \left(\left(\frac{11 N_c}{27}+1\right)f+\frac{1}{6}\right)g'^3
=\frac{41}{6}g'^3\\
\beta_\lambda &=& -3(3g^2+g'^2)\lambda+12\lambda^2 +
\frac{3}{4}(3 g^4+2 g^2 g'^2 + g'^4)+4 N_c\lambda y^2_t-4 N_c y^4_t\\
\beta_{\mu^2} &=&
\mu^2\left(-\frac{9}{2}g^2-\frac{3}{2}g'^2 + 6\lambda + 2 N_c y^2_t\right)
\end{eqnarray}

\subsection{Electroweak sector -- fermionic part}

We next turn to the fermionic part of the electroweak sector
given in (\ref{gammabargamma}). Taking the SM limit,
the one-loop divergent terms from this sector become 
\begin{eqnarray}
&& 32\pi^2\varepsilon\, {\cal L}^{\rm SM}_{\rm div,EWf} =\nonumber\\
&& \bar\psi_L\left(\frac{3}{2}g^2+2g'^2 Y^2_L\right)i\!\not\!\! D \psi_L
+\bar\psi_R\, 2g'^2 Y^2_R i\!\not\!\! D \psi_R
-8 g'^2\left(\bar\psi_L U Y_L{\cal M} Y_R \psi_R + {\rm h.c.}\right)\nonumber\\
&&+\frac{2}{v^2}
\bar\psi_L\langle{\cal M}_0{\cal M}^\dagger_0\rangle_I\, i\!\not\!\! D \psi_L
+\frac{4}{v^2}\bar\psi_R{\cal M}^\dagger_0{\cal M}_0\, i\!\not\!\! D \psi_R
\nonumber\\
&&-\frac{4}{v^2}\left(
\bar\psi_L U (\langle{\cal M}_0{\cal M}^\dagger_0\rangle_I -
{\cal M}_0{\cal M}^\dagger_0){\cal M}\psi_R +{\rm h.c.}\right)
\label{lsmdivewf}
\end{eqnarray}
Here $\langle\ldots\rangle_I$ denotes a trace over isospin indices only. 

Again, the nonrenormalizable operators in (\ref{gammabargamma}) have 
disappeared in the SM limit (\ref{lsmdivewf}).
The remaining divergences renormalize the fermionic part
of the SM Lagrangian, which can be written as
\begin{equation}\label{lsmf}
{\cal L}^{\rm SM}_{\rm f}=\bar\psi_L i\!\not\!\! D\psi_L
+ \bar\psi_R i\!\not\!\! D\psi_R 
-\left(\bar\psi_L(\tilde\Phi,\Phi){\cal Y}\psi_R +{\rm h.c.}\right)
\end{equation} 
We take
\begin{equation}\label{zpsiy}
(\tilde\Phi,\Phi)_{(0)} = Z^{1/2}_\Phi (\tilde\Phi,\Phi)\, ,\qquad
\psi^{(0)}_{L,R}=Z_{L,R}\psi_{L,R}\, ,\qquad 
{\cal Y}_{(0)}=\nu^\varepsilon ({\cal Y}+\Delta{\cal Y})
\end{equation}
where $Z_{L,R}=Z^\dagger_{L,R}$ are flavor matrices, which can be chosen
to be hermitean. $Z_\Phi$ is determined from the bosonic sector
discussed above.
From the definition of $\Delta{\cal Y}$ we find for the running of the Yukawa
matrix 
\begin{equation}\label{yrun}
\frac{d{\cal Y}}{dt}=\frac{1}{16\pi^2}\beta_{\cal Y}=
-\varepsilon {\cal Y} -\varepsilon\, \Delta{\cal Y}
-\frac{d}{dt}\Delta{\cal Y} +{\cal O}\left(\frac{1}{(16\pi^2)^2}\right)
\end{equation}

Inserting (\ref{zpsiy}) into the (unrenormalized) Lagrangian (\ref{lsmf}),
and using (\ref{ymrel}), (\ref{umphiy}) and $\Delta Z_X=Z_X - 1$, 
we find the counterterms
\begin{eqnarray}\label{lsmctewf}
{\cal L}^{\rm SM}_{\rm CT,EWf} &=&  \bar\psi_L\, 2\Delta Z_L\, i\!\not\!\! D\psi_L
+ \bar\psi_R\, 2\Delta Z_R\, i\!\not\!\! D\psi_R\nonumber\\
&-&(1+\eta)\left(\bar\psi_L U\left[ \Delta Z_L{\cal M}_0+{\cal M}_0\Delta Z_R+
\frac{1}{2}\Delta Z_\Phi {\cal M}_0 +\Delta{\cal M}_0\right]\psi_R 
+{\rm h.c.}\right) 
\end{eqnarray}
Requiring (\ref{lsmctewf}) to cancel the divergences 
of ${\cal L}^{\rm SM}_{\rm div,EWf}$ in (\ref{lsmdivewf}), we 
obtain $\Delta Z_{L,R}$ and $\Delta {\cal M}_0\equiv v\Delta{\cal Y}/\sqrt{2}$.
We find
\begin{equation}\label{dely}
\Delta{\cal Y}=-\frac{1}{32\pi^2\varepsilon}
\left[\left(\frac{9}{4}g^2 +\left(\frac{3}{4}+6 Y_L Y_R\right)g'^2
-N_c y^2_t\right){\cal Y}+
\frac{3}{2}\left(\langle{\cal Y}{\cal Y}^\dagger\rangle_I
-2{\cal Y}{\cal Y}^\dagger\right){\cal Y}\right] 
\end{equation}
From (\ref{yrun}) the beta function becomes
\begin{equation}\label{betay}
\beta_{\cal Y}=\frac{3}{2}\left(2{\cal Y}{\cal Y}^\dagger  
-\langle{\cal Y}{\cal Y}^\dagger\rangle_I \right){\cal Y}
-\left(\frac{9}{4}g^2 +\left(\frac{3}{4}+6 Y_L Y_R\right)g'^2
-N_c y^2_t\right){\cal Y}
\end{equation}
with $3/4+6 Y_L Y_R={\rm diag}(17/12,5/12,3/4,15/4)$,
in agreement with \cite{Celis:2017hod}.

\section{Conclusions}
\label{sec:concl}

The main results of this paper are:

\begin{itemize}

\item We computed for the first time the complete one-loop renormalization 
of the electroweak chiral Lagrangian including a light Higgs. All the 
divergent structures that we found either renormalize the LO Lagrangian or 
correspond to counterterms of the NLO Lagrangian according to the chiral 
counting of \cite{Buchalla:2013rka}. This result therefore corroborates that 
the chiral counting proposed in \cite{Buchalla:2013rka,Buchalla:2013eza} 
governs the divergence structure of the electroweak chiral Lagrangian.

\item We used the background-field method~\cite{Abbott:1981ke} to ensure 
explicit gauge invariance of background fields in all steps of the 
computation.  The divergent contributions to the one-loop effective Lagrangian 
were extracted using the super-heat-kernel~formalism~\cite{Neufeld:1998js}. 
As intermediate result, we rederived the 't Hooft master 
formula~\cite{tHooft:1973bhk}.

\item To cross-check the full result among ourselves, we have carried out 
independent calculations using different gauge fixing terms. We find full 
agreement in the final result.

\item We considered the SM limit as explicit cross-check of our result. 
We reproduce all the one-loop beta-functions in this limit. Considering only 
scalar (Goldstone and Higgs) fluctuations, we further reproduce 
\cite{Guo:2015isa}, which was also cross-checked later 
by \cite{Alonso:2015fsp}.

\end{itemize}

\section*{Note added}

After the present paper had been made public on arXiv, the article
\cite{Alonso:2017tdy} appeared, in which essentially the same topic
is addressed, and which includes the formulation of renormalization-group
equations.

\section*{Acknowledgements}
C. K. and M. K. wish to thank the Ludwig-Maximilian University
for the warm hospitality extended to them during the course of this work.
This work was supported in part by the DFG cluster of excellence EXC 153
`Origin and Structure of the Universe', DFG grant BU 1391/2-1,
DFG grant FOR 1873, the Bundesministerium f\"ur Bildung und Forschung
(BMBF FSP-105), 
the Spanish Government and ERDF 
funds from the EU Commission [Grants No. FPA2014-53631-C2-1-P and 
SEV-2014-0398]. A.C. was supported in part
by a Research Fellowship of the Alexander von Humboldt Foundation.
The work of M. K. has been carried out thanks to the support of the 
OCEVU Labex (ANR-11-LABX-0060) and the A*MIDEX project (ANR-11-IDEX-0001-02) 
funded by the "Investissements d'Avenir" French government 
program managed by the ANR.


\end{document}